\documentclass[fleqn,twoside]{article}
\usepackage{espcrc2}
\usepackage{graphicx}
\usepackage[figuresright]{rotating}

\catcode`@=11
\def\gsim{\mathrel{\mathpalette\@versim>}}
 \def\@versim#1#2{\lower0.2ex\vbox{\baselineskip\z@skip\lineskip\z@skip
       \lineskiplimit\z@\ialign{$\m@th#1\hfil##$\crcr#2\crcr\sim\crcr}}}
\catcode`@=12 

\newcommand{\AmS}{{\protect\the\textfont2
  A\kern-.1667em\lower.5ex\hbox{M}\kern-.125emS}}

\title{Tests of the Standard Model with Low--Energy Neutrino Beams}

\author{V.~Antonelli\address[Fisica]{Dipartimento di Fisica, Universit\`a, 
di Milano, Via Celoria 16, I-20133 Milano,
Italy}\address[INFN]{I.N.F.N., Sezione di Milano, Via Celoria 16, I-20133
    Milano, Italy}, G. Battistoni\addressmark[INFN],
  P. Ferrario\address{IFIC, CSIC-Universitat de Val\`encia, Apt.  Correos 
22085, E-46071
Valencia, Spain} and S. Forte\addressmark[Fisica]\addressmark[INFN].}

\begin{document}

\begin{abstract}

We discuss the possibility of using future high--intensity low--energy
neutrino beams  
for precision tests of the Standard Model. In particular we consider the 
determination of the electroweak mixing angle from elastic and
quasi--elastic neutrino--nucleon scattering at a superbeam or $\beta$--beam.


\end{abstract}

\maketitle
\section{The High--Intensity Frontier and Neutrino Beams}

The search for new physics beyond the standard model in the future
will follow two main paths: higher energy and
high--intensity~\cite{foster}.
Neutrino facilities will play an important role in this program: 
neutrino masses take us beyond the standard
model, and a full determination of the pattern of $\nu$ masses and
mixing will require dedicated high--intensity neutrino
beams~\cite{blondel}.
However, a high--intensity $\nu$ beam can be used  not 
only to study neutrino properties, but also as 
a sensitive probe of the electroweak
interaction. Indeed, it
has been shown that a wide spectrum of otherwise
very difficult or impossible measurements of strong and electroweak
processes would become possible at a high--energy
neutrino factory~\cite{nufact}. 

Here we start addressing the issue whether equally interesting
measurements might also be possible with a {\it low energy} but
sufficiently intense $\nu$ beam. Indeed, the main role so far of
low--energy tests of the standard model~\cite{musolf} has been in the
study of rare processes.  Here we will investigate whether with
sufficiently high--intensity they may lead to competitive measurements
of standard model parameters, specifically the electroweak mixing angle.

\section{Future Perspectives of Neutrino Physics}
\vspace{-0.2 truecm}
The development of future neutrino facilities is driven by the study
of neutrino masses and mixings, and will happen in three stages.
In the first phase various facilities
will produce conventional $\nu$
beams from the decay of  a secondary meson beam (producing typically
$\sim10^{18}$ $\nu$/year). 
Examples of such facilities (now under construction or just
commissioned) are MINOS and NO$\nu$A, the CERN/Gran Sasso beam, and  T2K.
In the second stage, planned for the beginning of the next decade,
``superbeams'', i.e. conventional beams but with intensities about 
hundred times higher, should be constructed, exploiting very high--intensity, 
and relatively low energy primary proton beams. Examples of such facilities 
are the second phase of T2K, exploiting  a 50 GeV and 7 MW proton synchrotron 
at JParc, and a possible high--energy superbeam at CERN exploiting  a
3.5~GeV, 4--5~MW superconducting proton linac (SPL).
In the third phase, starting perhaps towards the end of the next
decade, neutrinos from decays of a {\it primary} beam are planned. Two
possible kinds of primary beams are envisaged: either a neutrino factory,
i.e. relatively high--energy
(tens of GeV) muon beam, or a $\beta$--beam, i.e.  relatively low--energy (few
GeV) radioactive nuclei. The advantage of using a primary beam are
higher intensity and a better control on the neutrino energy spectrum.
The $\beta$ beam has the further advantage of giving a pure beam of electronic
$\nu$ or $\bar{\nu}$, with essentially no contamination from $\nu_{\mu}$ or
wrong--sign neutrinos. The neutrino factory gives a beam which is exactly 50\% $\nu_\mu$ and $\bar \nu_e$.

\section{The Weinberg Angle from Neutrino-Nucleon Elastic Scattering} 

Because neutrinos only couple to weak interactions, they are an
ideal probe of electroweak parameters, specifically of the electroweak
mixing angle (Weinberg angle) which controls the relative strength of
neutral (NC)  and charged current (CC) couplings. Neutrino--electron
elastic scattering offers an ideally clean setting for this
measurement, which is competitive at a high energy neutrino 
factory~\cite{nufact}, but (because the cross
section grows linearly with the energy) at a low--energy
facility very high intensities are required~\cite{leel}.
A measurement of the Weinberg angle can be obtained from the
CC/NC deep-inelastic scattering ratio: this measurement at present is almost
competitive, but marred by the uncertainty related to parton
distributions~\cite{nutev}. It would certainly be competitive at a
neutrino factory~\cite{nufact}. 

\begin{figure}[htb]
\vspace{-0.9cm}
\includegraphics[width=15pc]{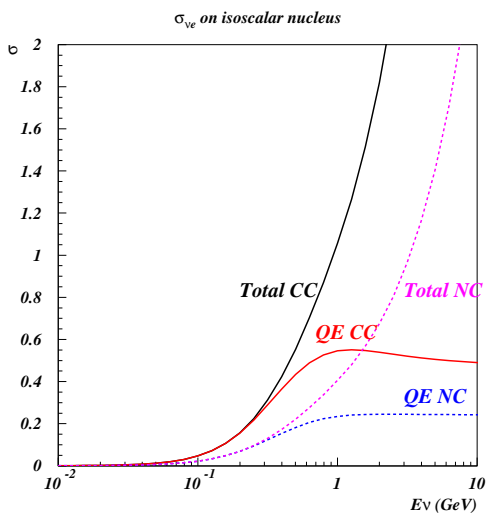}
\vspace{-1.0cm}
\caption{$\nu$--nucleon cross sections vs. energy.}
\vspace{-0.9cm}
\label{xsect}
\end{figure}

As the energy is lowered, the relative 
(quasi)elastic contribution to the total cross section grows, and
at an energy $E\sim M_N$ the elastic and inelastic contribution are of
comparable size (see  Fig.~\ref{xsect}). 
At this energy, at which perturbative
treatment of inelastic contribution breaks down anyway, the
relative elastic contribution is sizable, while total cross
section is still reasonably large. This energy is relevant for
future facilities, such as JPARC, a low--energy $\beta$ beam
or the SPL superbeam, and it is natural to ask whether elastic or
quasielastic scattering can be used for competitive measurements of
Weinberg angle.

The answer is not obvious, because (quasi)--elastic 
cross sections depend on eight independent form
factors~\cite{alberico}:  two pairs of electric and magnetic form
factors for proton and neutron targets, a pair of strange electric and
magnetic form factors (the same for protons and neutrons) and a pair
of axial isotriplet and strange form factors (the same for protons and
neutrons, up to signs). 
The question is then whether the uncertainty in the form
factors knowledge spoils the extraction of $\sin^2\theta_W$ from these
cross sections.

\section{Results}

\subsection{Physical observables}
The simplest answer to the above question is obtained by a counting 
of the relevant physical observables. With proton and
neutron (from deuterium or other nuclei) targets and $\nu$ and $\bar \nu$
beams, one can measure four independent NC and two independent CC
cross sections.  They depend on
eight form factors and the Weinberg angle. Hence, at least three form
factors have to be input to the analysis. It is convenient to
input the electric form factors, whose forward value is fixed by
charge. 

Assuming a flux $\Phi_\nu\sim\Phi_{\bar{\nu}}\sim
10^{11}/(m^2 yr)$ with energy $
  E_\nu\sim
  E_{\bar \nu}\approx 5$~GeV one gets 
$\sim 10^5$ elastic CC events and  $\sim
  10^4$ NC events with either beam or target after one year of running
  with each beam. This flux and energy are typical e.g. of a low--energy
  $\beta$--beam, with a detector located  at a distance of $\sim100$~Km. 
 Assuming that the five
  independent form factors and the Weinberg angle are determined in
  each angular bin this leads to a statistical error $\Delta
  \sin^2\theta_w\sim 10^{-3}$~\cite{ferrario}. 

The further theoretical 
  error due to the electric form factor is negligible.  It is
  important to observe that these form factors must be input if one
  wishes to extract all the other form factors and 
 $\sin^2\theta_w$. However, the cross section can actually be
 measured in a large number of angular (or $y$) bins (e.g. several dozens). One may
 thus choose to parametrize the form factors and fit these parameters
 as well as $\sin^2\theta_w$. Clearly, with, say, 20 bins and several
 cross sections even with a very general parametrization all form
 factors can be determined together with the Weinberg angle.
This suggests  that a more detailed analysis is worthwhile. 

\subsection{Experimental  constraints}

The main experimental constraint is the possibility to detect CC and
NC events with a neutrino beam energy between one and a few GeV. This
rules out  water Cherenkov 
detectors, because the Cherenkov threshold $\frac{p}{E}>0.75$ for the
recoiling proton implies that only protons with recoil momentum $p>1.1$~GeV
can be detected, which removes most of the cross section. A more
promising alternative is a liquid Ar TPC~\cite{icarus}. In this case, the only
constraint is that the recoiling proton leaves a sufficiently long
track so that it is not confused with nucleon motion due to nuclear
effects. This gives a constraint on the proton energy $E-m\sim50$~MeV
i.e. $p\gsim300$~MeV. With a beam energy of the order of 1 GeV, about 75\% of the
scattering events survives this kinematic cut. However, recoiling
neutrons cannot be detected. This implies that neutron neutral current (NC)
events are essentially lost and one is left with only four independent cross sections. 

In order to maximize the flux, one may envisage the option of having a
near detector, located at a few hundreds of meters from the source,
thereby obtaining fluxes by many orders of magnitude larger than those
at the far detector used for oscillations studies. However,
in a realistic analysis one should keep into account that
an Argon TPC might have difficulties in  handling  interaction rates
much larger than a few
events per spill. This would put a bound on the maximum flux.

\subsection{Quantitative analysis}

In order to get a more quantitative estimate of the accuracy one can reach in 
the Weinberg angle determination, we have generated scattering
events assuming an incoming flux 
$\Phi_\nu \simeq 10^{16}/(m^2 yr)$, $\Phi_{\bar \nu} \simeq 5\times 10^{14}/(m^2 yr)$,
with fixed energy $E=1$~Gev. These parameters are typical e.g. of  T2K
(first phase) with a near detector at about 300~m from the source. In the 
present analysis we have considered the case of a liquid Argon detector 
with a mass of 10~KTons. An increase in the detector mass would correspond to a
reduction in the fit uncertainty that can be easily obtained by standard 
statistical analysis.
We have assumed $\sin^2\theta_w=0.2312$ and all the nucleon form factors as 
given in ref.~\cite{ff} and in ref.~\cite{garvey,alberico} for the axial and 
strange form factors; in particular for the strange magnetic 
form factor we used $G_M^S(Q^2) = \frac{F_1^S Q^2 + F_2^S (0)}{\left(1+
    \tau\right) \left(1+\frac{Q^2}{M_V^2}\right)^2}$, with
$Q^2 = - (k-k^{'})^{2}$ the neutrino momentum transfer, 
$\tau= Q^2/(4 M_N^2)$, $M_N$ nucleon mass, $M_V=0.843 \, \rm{GeV/c}^2$ 
and $F_1^S=0.49$. 

We have then performed a fit to the events thus generated,
 leaving as free parameters 
$\sin^2\theta_w$ and the forward value of the strange magnetic form factor 
$G_M^S(0)$. We have then repeated this fit by varying the value of the 
forward strange axial form factor $G_A^S(0)$, considering one $\sigma$
variation around its central value ($G_A^S(0)=-0.13 \pm 0.09$). 
These are the only forward form factors which are affected by a significant
uncertainty. Other form factor parameters have a more moderate impact.  
We get
$\sin^2\theta_w=0.2309\pm0.0019\hbox{(stat)}\pm0.0024\hbox{(syst)}$, 
where the systematic error is due to the variation of the strange
axial form factor within the range indicated. 
Clearly, a more detailed analysis~\cite{noibeam} would require either fitting
of all form factors, or varying some of their parameters within
errors. In such an analysis we will also introduce a study of the systematical
uncertainty related to the choice of the form factor parametrization.
However, on the basis of this first estimates, we conclude that
a determination of $\sin^2\theta_w$  with an uncertainty of a few percent
is not unreasonable. 

\section{Acknowledgments}

One of us, V.A., is deeply grateful to the NOW 2006 organizers for providing 
a stimulating human and scientific atmosphere. He also would like to thank 
in particular A. Guglielmi, M. Mezzetto and E. Torrente-Lujan for useful 
discussions.

\end{document}